\documentclass[useAMS,usenatbib,a4paper]{mn2e}

\usepackage{epsfig,bm}
\usepackage{amsbsy,amssymb,amsmath}
\usepackage{hyperref}
\usepackage{graphicx}
\usepackage{dcolumn}
\usepackage{ctable}
\usepackage{natbib}
\usepackage{float}
\usepackage{color}

\newcommand{\be}{\begin{equation}}
\newcommand{\ee}{\end{equation}}
\newcommand{\bea}{\begin{eqnarray}}
\newcommand{\eea}{\end{eqnarray}}

\newcommand{\bl}{\mathbf{l}}
\newcommand{\kpe}{k_\perp}
\newcommand{\kpa}{k_\parallel}

\title[Testing gravity with HI intensity mapping]
{Testing gravity at large scales with HI intensity mapping}
\author[ Pourtsidou]{ Alkistis Pourtsidou  \\ 
Institute of Cosmology \& Gravitation, University of Portsmouth, Burnaby Road, Portsmouth, PO1 3FX, United Kingdom 
}

\begin{document}

\maketitle

\begin{abstract}
We investigate the possibility of testing Einstein's general theory of relativity (GR) and the standard cosmological model via the $E_{\rm G}$ statistic using neutral hydrogen (HI) intensity mapping. We generalise the Fourier space estimator for $E_{\rm G}$ to include HI as a biased tracer of matter and forecast statistical errors using HI clustering and lensing surveys that can be performed in the near future, in combination with ongoing and forthcoming optical galaxy and Cosmic Microwave Background (CMB) surveys. We find that fractional errors $< 1\%$ in the $E_{\rm G}$ measurement can be achieved in a number of cases and compare the ability of various survey combinations to differentiate between GR and specific modified gravity models. Measuring $E_{\rm G}$ with intensity mapping and the Square Kilometre Array can provide exquisite tests of gravity at cosmological scales. 
\end{abstract}


\section{Introduction}

During the last two decades, observational cosmology has entered an era of unprecedented precision. The standard cosmological model ($\Lambda$CDM) fits the data extremely well, but requires that General Relativity (GR) is the correct description of gravity on all scales and that the matter-density of our Universe is dominated by the two constituents of the dark sector, {\it i.e.} dark energy in the form of a cosmological constant and cold dark matter. Dark energy is thought to be responsible for the accelerated expansion of the Universe \citep{Riess:1998cb,Perlmutter:1998np} and uncovering its nature is arguably the most exciting challenge in modern cosmology. Alternative explanations to the cosmological constant have been proposed, for example a dynamically evolving scalar field playing the role of dark energy (see \citet{Copeland:2006wr} for a review).  A different point of view suggests that late time cosmic acceleration could be due to modifications to the laws of gravity on the largest (cosmological) scales (see \citet{Clifton:2011jh} for a review). In general, exotic dark energy and modified gravity theories modify the background and perturbation evolution and dynamics of the Universe and they have distinct and detectable observational effects.

In this paper we are going to investigate the possibility of testing $\Lambda$CDM and the laws of gravity at large scales using the $E_{\rm G}$ statistic, which was first introduced in \citet{Zhang:2007nk}. 
The definition of $E_{\rm G}$ in Fourier space is
\be
E_{\rm G}(k,z) = \frac{c^2k^2(\phi-\psi)}{3H^2_0(1+z)\theta(k)} \, ,
\label{eq:EG}
\ee where $(\phi,\psi)$ are the scalar potentials in the perturbed Friedmann-Robertson-Walker (FRW) metric $ds^2=(1+2\psi)dt^2-a^2(1+2\phi)d\mathbf{x}^2$, $\theta \equiv \mathbf{\nabla \cdot v}/H(z)$ is the peculiar velocity perturbation field, and $H_0$ is the value of the Hubble parameter today. 

From the above definition it is clear that $E_{\rm G}$ depends on how gravity behaves on large scales.  In GR, assuming that the background Universe is described by a flat FRW metric and in the absence of anisotropic stress, we can show that the Poisson and anisotropy equations can be written as 
\begin{align} \nonumber
\centering
k^2\psi &= -4\pi Ga^2\rho \delta \\
\phi &= -\psi \, ,
\end{align} where $a$ is the scale factor, $\rho$ is the background matter density and $\delta$ is the matter density perturbation. In modified gravity (MG) we can use two scale- and time-dependent functions $\mu(k,a)$ and $\gamma(k,a)$ to parametrise possible departures from $\Lambda{\rm CDM}$ and write \citep{Hojjati:2011ix}
\begin{align} \nonumber
\centering
k^2\psi &= -4\pi Ga^2\mu(k,a)\rho \delta \\
\phi &= -\gamma(k,a)\psi \, .
\end{align} 
Substituting these expressions in the $E_{\rm G}$ definition (\ref{eq:EG}) and using the fact that on linear scales we can write $\theta = f\delta$, where $f$ is the linear growth rate, we find  \citep{Pullen:2015a} 
\be
E_{\rm G}(k,z)=\frac{\Omega_{{\rm m,0}}\mu(k,a)[1+\gamma(k,a)]}{2f} \, .
\label{eq:EGmugamma}
\ee For GR, we set $\mu=\gamma=1$ and find $E_{\rm G}(k,z)=\Omega_{{\rm m,0}}/f(z)$, with 
 $\Omega_{{\rm m,0}}$ the matter density today relative to the critical density.

In order to measure $E_{\rm G}$, the estimator in \citet{Zhang:2007nk} involved the ratio between the cross-correlation power spectrum of galaxies and weak lensing convergence $P_{\rm g\kappa}$ and the cross-correlation power spectrum of galaxies and velocities $P_{\rm g \theta}$. The latter is equivalent to the galaxy autocorrelation times the redshift-space distortion (RSD) parameter $\beta$, {\it i.e.} $\beta P_{\rm gg}$, where $\beta = f/b_g$ with $b_g$ the galaxy clustering bias. In \citet{Reyes:2010tr} and \cite{Blake:2015vea} $E_{\rm G}$ was measured using a real space estimator and the results were fully consistent with GR. Recently, forecasts for future galaxy surveys were presented in \citet{Leonard:2015cba}.

In \citet{Pullen:2015a} a Fourier estimator for $E_{\rm G}$ was constructed in terms of the galaxy-CMB lensing and galaxy clustering angular power spectra, also including the RSD parameter $\beta$. Very recently, they reported the largest-scale measurement of $E_{\rm G}$ using the aforementioned method; they found $E_{\rm G}(z=0.57)=0.243\pm 0.060 \,\, {\rm (stat)} \pm 0.013 \,\, {\rm (sys)}$, a result in $2.6\sigma$ tension with the GR prediction at this redshift \citep{Pullen:2015b}.

In the next Section we will closely follow the approach by \citet{Pullen:2015a} and generalise their $E_{\rm G}$ Fourier space estimator to include a different dark matter tracer than galaxies, namely neutral hydrogen (HI). In Section~\ref{sec:surveys} we describe the various optical galaxy, intensity mapping and CMB surveys we are going to use to forecast future $E_{\rm G}$ measurements. We present our forecasts in Section~\ref{sec:forecasts} and demonstrate how they can be used to discriminate between GR and specific modified gravity models. We conclude in Section~\ref{sec:conclusions}.
 
\section{Formalism}
\label{sec:formalism}

The Fourier space estimator for $E_{\rm G}$ is constructed as \citep{Pullen:2015a} 
\be
\hat{E}_{\rm G}(\ell,\bar{z})=\frac{c^2\hat{C}_\ell^{\rm g\kappa}}{3H^2_0\hat{C}_\ell^{\rm g\theta}} \, ,
\ee and it can be expressed in terms of the
 galaxy-convergence angular cross-power spectrum $C^{\rm g\kappa}_\ell$, the galaxy angular auto-power spectrum $C^{\rm gg}_\ell$, and the RSD parameter $\beta = f/b_g$ (see also \citet{Reyes:2010tr}). Note that it is galaxy bias free in the linear regime, which evades the issue of possible degeneracies between modified gravity effects and bias.

In this work, we are going to utilize the above estimator but using HI instead of galaxies. We will forecast statistical errors for $E_{\rm G}$ using HI intensity mapping (IM) clustering and lensing surveys which can be performed in the near future using the Square Kilometre Array (SKA)\footnote{www.skatelescope.org}. Intensity mapping \citep{Battye:2004re,Chang:2007xk,Loeb:2008hg,Mao:2008ug,Peterson:2009ka,Seo:2009fq,Ansari:2011bv,Battye:2012tg,Switzer:2013ewa,Bull:2014rha} is an innovative technique which uses HI to map the large-scale structure of the Universe in three dimensions. Instead of detecting individual galaxies like the conventional galaxy surveys, intensity mapping surveys use HI as a dark matter tracer by measuring the intensity of the redshifted 21cm line across the sky and along redshift, treating the 21cm sky as a diffuse background. 

Using HI instead of galaxies we can write the Fourier space $E_{\rm G}$ estimator as
\be
\hat{E}_{\rm G}(\ell,\bar{z})=\frac{c^2\hat{C}_\ell^{{\rm \delta_{\rm HI}{\rm \kappa}}}}{3H^2_0\hat{C}_\ell^{{\rm \delta_{\rm HI}}{\rm \theta}}} \, ,
\label{eq:EGest}
\ee 
where $\delta_{\rm HI}$ is the HI density contrast which traces the matter density as $\delta_{\rm HI}=b_{\rm HI}\delta$, with $b_{\rm HI}$ the HI bias.
The HI-convergence angular cross-power spectrum using the Limber approximation \citep{Limber:1954zz} for scales $\ell \geq 10$ is given by
\be
C_\ell^{\delta_{\rm HI}\kappa}=\frac{3\Omega_{{\rm m},0}H^2_0}{2c^2}b_{\rm HI}(\bar{\chi})(1+\bar{z})P_{\delta \delta}(\ell/\bar{\chi},\bar{\chi}) \bar{\chi}^{-1}(1-\bar{\chi}/\chi_s) \, ,
\ee where $\bar{\chi}$ is the comoving radial distance to redshift $\bar{z}$, $\chi_s$ is the comoving radial distance to the background sources, and $P_{\delta \delta}$ is the matter power spectrum. 
When deriving this formula we assumed that both the foreground lenses and the background sources distributions can be approximated by delta functions. This is going to be a valid assumption in the cases we are going to study, as our chosen foreground redshift bins are always going to be small enough ($\Delta z = 0.1$), and for the background sources we use either the CMB plane or 21cm sources, both of which can be well approximated by a delta function distribution. 

Then, following the formalism \citet{Pullen:2015a} used for the galaxy clustering case, we construct the velocity-HI angular cross-power spectrum, to be
\be
C^{\delta_{\rm HI} \theta}_\ell = \frac{\bar{\chi}(1-\bar{\chi}/\chi_s)\beta(\bar{z})(1+\bar{z})}{2\Delta \chi^{-1}}\hat{C}^{\delta_{\rm HI}\delta_{\rm HI}}_\ell \, ,
\ee with $\Delta \chi$ the comoving width of the redshift bin with central redshift $\bar{z}$, and
\be
C^{\delta_{\rm HI}\delta_{\rm HI}}_\ell =  \frac{1}{\Delta \chi} \bar{\chi}^{-2}b^2_{\rm HI}(\bar{\chi})
P_{\delta \delta}(\ell/\bar{\chi},\bar{\chi}) \, .
\ee
Note that the RSD parameter is $\beta=f/b_{\rm HI}$. 

The fractional error of $E_{\rm G}$ can be written as
\be
\frac{\sigma^2[E_{\rm G}(\ell,\bar{z})]}{E^2_{\rm G}}=\left[\left(\frac{\sigma(C_\ell^{{\delta_{\rm HI}}{\rm \kappa}})}{C_\ell^{{\delta_{\rm HI}}{\rm \kappa}}}\right)^2+\left(\frac{\sigma(\beta)}{\beta}\right)^2
+\left(\frac{\sigma(C_\ell^{{\delta_{\rm HI}}{\delta_{\rm HI}}})}{C_\ell^{{\delta_{\rm HI}}{\delta_{\rm HI}}}}\right)^2\right].
\label{eq:EGerror}
\ee
The error in the measurement of $C_\ell^{\rm \delta_{\rm HI}{\rm \kappa}}$ is
\be
\sigma^2(C_\ell^{\rm \delta_{\rm HI}{\rm \kappa}})=
\frac{(C_\ell^{\rm \delta_{\rm HI}{\rm \kappa}})^2+(C_\ell^{\rm \kappa \kappa}+N_\ell^{\rm \kappa \kappa})(C_\ell^{\delta_{\rm HI}\delta_{\rm HI}}+N_\ell^{\delta_{\rm HI}\delta_{\rm HI}})}{(2\ell+1)f_{\rm sky}} \, ,
\ee where $C_\ell^{{\rm \kappa \kappa}}$ is the lensing convergence power spectrum, $N_\ell^{{\rm \kappa \kappa}}$ the lensing reconstruction noise, $N_\ell^{\delta_{\rm HI}\delta_{\rm HI}}$ the noise in the measurement of the HI clustering angular power spectrum $C_\ell^{\delta_{\rm HI}\delta_{\rm HI}}$, and $f_{\rm sky}$ the (overlapping) fraction of the sky scanned by the surveys used.

Note that in the following we are also going to consider combinations of galaxy-CMB lensing surveys, like the ones presented in \citet{Pullen:2015a}. The formulae for this case can be recovered by setting $\delta_{\rm HI}\rightarrow \delta_g$ and $b_{\rm HI}\rightarrow b_g$ in the above equations. 
We should also mention that Eq.~(\ref{eq:EGerror}) is exact only if $\beta$ and $C_\ell^{\delta_{\rm HI}\delta_{\rm HI}}$ (or $C_\ell^{gg}$ in the optical galaxy case) are measured by different, independent surveys --- otherwise they should be correlated. However, we can safely use Eq.~(\ref{eq:EGerror}) for the cases we are going to consider in this paper. That is because the dominant errors are the lensing and $\beta$ ones, and the clustering error due to thermal noise (shot noise) for the upcoming IM (photometric galaxy) surveys we will use is very small so it can be neglected. Thus, the combined ($\beta C^{{\rm HI}-{\rm HI}}$) error increases only slightly via the covariance and our results are practically unchanged (see \citet{Pullen:2015a} for a detailed discussion of this issue for the optical galaxy case).

Before we move on to our forecasts, we will dedicate the next Section to the specifications and noise properties of the various surveys we are going to use.

\section{The surveys}
\label{sec:surveys}

\subsection{Galaxy and HI IM clustering}

For our forecasts involving galaxy clustering we will consider two photometric surveys: the ongoing Dark Energy Survey (DES)\footnote{http://www.darkenergysurvey.org/} and the planned Large Synoptic Survey Telescope (LSST)\footnote{http://www.lsst.org/}. Both of them aim to investigate the nature of cosmic acceleration and are able to perform precision galaxy clustering measurements. 

The DES survey parameters are $A_{\rm sky}=5000 \, {\rm deg}^2$ (equivalently, $f_{\rm sky}\simeq 0.1$), number density of galaxies $n_g = 10 \, {\rm arcmin}^{-2}$, and redshift range $0<z<2$ with median redshift $z_0=0.7$ \citep{Becker:2015ilr}. 
The LSST survey parameters are assumed to be $f_{\rm sky}=0.5$, number density of galaxies $n_g = 40 \, {\rm arcmin}^{-2}$, and redshift range $0<z<2.5$ with median redshift $z_0=1$ \citep{Abell:2009aa}. 
For our forecasts we model the redshift distribution of galaxies as
\be
\frac{dn}{dz} \propto z^2 \, {\rm exp}[-(z/z_0)^{3/2}]
\ee and use $b_g(z)=\sqrt{1+z}$ for the galaxy clustering bias. The noise term $N_\ell^{\delta_g\delta_ g}=1/\bar{n}_i$, where $\bar{n}_i$ is the number of galaxies per steradian in the $i$-th redshift bin.

For HI clustering using the intensity mapping technique we will consider the SKA\_Mid instrument. This can operate in two observing modes, the single-dish (autocorrelation) mode and the interferometer mode. Their noise properties have been described in detail in \citet{Bull:2014rha,Pourtsidou:2015mia}, so we are just going to state the relevant formulae here.  
The thermal noise angular power spectrum for the single-dish mode is given by
\be
C_\ell^{\rm N} = \Omega_{\rm pix}(\sigma_{\rm pix})^2{\rm exp}[\ell(\ell+1)(\theta_{\rm B}/\sqrt{8{\rm ln}2})^2],
\ee where the pixel thermal noise is $\sigma_{\rm pix}=T_{\rm sys}/\sqrt{2Bt_{\rm obs}}$, and
$\Omega_{\rm pix}\simeq 1.13\theta^2_{\rm B}$. $T_{\rm sys}$ is the system temperature, $B$ the bandwidth of observation, $t_{\rm obs}$ the observation time, and $\theta_{\rm B} \sim \lambda/D_{\rm dish}$ the beam FWHM of a dish with diameter $D_{\rm dish}$ at wavelength $\lambda$.
The observation time is given by $t_{\rm obs}=t_{\rm tot}(\Omega_{\rm pix}/\Omega_{\rm tot})N_{\rm dishes}$, where $t_{\rm tot}$ is the total survey time, $\Omega_{\rm tot}$ is the sky area the survey scans, and $N_{\rm dishes}$ the number of available dishes. For the interferometer mode the thermal noise power spectrum is given by \citep{Pourtsidou:2015mia}
\be
C_\ell^{\rm N} = \frac{T^2_{\rm sys}{\rm [FOV]^2}}{Bt_{\rm tot}n(\ell)}
\ee with ${\rm FOV} \simeq (\lambda/D_{\rm dish})^2$ and $n(\ell)$ the number density of baselines.

The HI clustering noise term will be calculated as 
\be
N^{\delta_{\rm HI}\delta_{\rm HI}}_\ell = \frac{C^{\rm N}_\ell}{\bar{T}^2},
\ee where the mean brightness temperature $\bar{T}(z)$ is given by
\be
\bar{T}(z)=180 \, \frac{\Omega_{\rm HI}(z)h(1+z)^2}{H(z)/H_0}  \, {\rm mK} \, .
\ee 
For our forecasts we will assume that the HI density evolves with redshift as
$
\Omega_{\rm HI}(z) = 4 \times 10^{-4} (1+z)^{0.6}
$ which has been suggested in \citet{Crighton:2015pza}.
We are also going to use the HI bias model $b_{\rm HI}(z)$ from \citet{Camera:2013kpa} and consider Phase 1 of the SKA\_Mid instrument (SKA1\_Mid), consisting of $130$ dishes with $15$ m diameter according to the recently updated specifications (`re-baselining') in order to meet budget constraints \citep{McPherson15}. The redshift range is $0.35<z<3$ (Band 1). The system temperature is given by \citep{Dewdney13}
\be
T_{\rm sys} = 28 + 66\left(\frac{\nu}{300 \, {\rm MHz}}\right)^{-2.55} \, {\rm K} \, ,
\ee with $\nu$ the observing frequency. The $n(\ell)$ distribution is taken from \citet{Bull:2014rha}. The bandwidth $B$ will be determined by the width of our chosen redshift bins, which is going to be $\Delta z=0.1$ for all cases. Finally, we will consider a survey strategy with $t_{\rm tot}=4,000 \, {\rm hrs}$ and $f_{\rm sky}=0.5$. An important point is the difference in the range of angular scales that the dish and interferometer mode can observe. We will analyse this in detail when we present our forecasts in Section~\ref{sec:forecasts}.

\subsection{CMB and 21cm lensing}

Following  \citet{Pullen:2015a} we will study the case where CMB is the background source plane for the lensing convergence measurements. We will calculate the lensing reconstruction noise using the formalism by \citet{Hu:2001kj}. The instrumental noise for a CMB survey is given by
\be
 C^{\rm N}_\ell = \Delta^2_T \, {\rm exp}[\ell(\ell+1)\sigma^2/8{\rm ln}2].
\ee 
We will consider a future COrE-like satellite with FWHM $\sigma = 3.0^\prime$ and temperature noise $\Delta_T = 1 {\, \rm \mu K^\prime}$. We will also consider the full \emph{Planck} lensing map ({\it i.e.} including temperature and polarization) using the sensitivities given by the \emph{Planck} Collaboration \citep{Planck:2006aa}.

For the 21cm lensing case we will use two different estimators depending on the chosen source redshift. The first one assumes the temperature distribution is Gaussian, which is a reasonable approximation at the Epoch of Reionization (EoR), at least while the ionised regions are small. This Fourier space estimator was developed in \citet{Zahn:2005ap} and it is a 3D extension of the CMB lensing estimator by \citet{Hu:2001kj}. More specifically, 
the 21cm brightness temperature fluctuations are divided into wave vectors perpendicular to the line of sight $\mathbf{\kpe}=\mathbf{l}/{\cal D}$, with ${\cal D}$ the angular diameter distance to the source redshift, and a
discretised version of the parallel wave vector $\kpa =\frac{2\pi}{{\cal L}}j$ where ${\cal L}$ is the depth of the observed volume. Considering modes with different $j$ as independent, an optimal
estimator can be found by combining the individual estimators for
different $j$ modes without mixing them. The three-dimensional lensing reconstruction noise is then found to be \citep{Zahn:2005ap}
\begin{align} \nonumber
&N^{\kappa \kappa}_\ell = (\ell^4/4) \times \\
&\left[\sum_{j=j_{\rm min}}^{j_{\rm max}} \int \frac{d^2\ell'}{(2\pi)^2}  \frac{[\bl' \cdot \bl C_{\ell',j}+\bl \cdot (\bl-\bl')
C_{|\ell'-\ell|,j}]^2}{2 C^{\rm tot}_{\ell',j}C^{\rm tot}_{|\ell'-\ell|,j}}\right]^{-1},
\label{eq:Nkkell}
\end{align}
where
\be \nonumber
\label{eq:Cellj}
C_{\ell,j} = [\bar{T}(z)]^2 \frac{P_{\delta \delta}(\sqrt{(\ell/{\cal D})^2+(j2\pi/{\cal L})^2})}{{\cal D}^2 {\cal L}}
\ee  and
\be \nonumber
C^{\rm tot}_{\ell, j} = C_{\ell,j} + C^{\rm N}_\ell.
\ee
For the EoR case we will use an SKA\_Low-like interferometer array with uniform antennae distribution for which the thermal noise power spectrum is given by
\citep{Zaldarriaga:2003du}
\begin{equation}
\label{eq:CellN}
C^{\rm N}_\ell = \frac{(2\pi)^3 T^2_{\rm sys}}{B t_{\rm tot} f^2_{\rm cover} \ell_{\rm max}(\nu)^2} \, ,
\end{equation} where  $\ell_{\rm max}(\nu)$ is the  highest multipole that can be measured by the array at frequency $\nu$ (wavelength $\lambda$) and is related to $D_{\rm tel}$, the maximum baseline of the core array, by $\ell_{\rm max}(\lambda)=2\pi D_{\rm tel}/\lambda$;
 $f_{\rm cover}$ is the total collecting area of the core array, $A_{\rm coll}$ divided by $\pi(D_{\rm tel}/2)^2$.

Finally, for post-reionization redshifts we will use SKA\_Mid in interferometer mode and the lensing estimator developed in \citet{Pourtsidou:2013hea,Pourtsidou:2014pra} using the intensity mapping technique. This estimator takes into account the discreteness of galaxies, and models the HI distribution as a Poisson distribution drawn from a Gaussian distribution that represents the clustering of galaxies. Further details and results for the lensing reconstruction using this technique and the updated SKA\_Mid instrument have been analysed in detail in recent work \citep{Pourtsidou:2015mia} so we will not repeat them here. We just note that $N^{\kappa \kappa}_\ell$ for this case involves the underlying dark matter power spectrum $P_{\delta \delta}$, the HI density $\Omega_{\rm HI}(z)$ as well as the HI mass (or luminosity) moments up to $4$th order and, of course, the thermal noise of the instrument.

\section{Forecasts}
\label{sec:forecasts}

In this Section we will present our forecasts for the constraining power of combinations of intensity mapping, optical galaxy, and CMB surveys using the $E_{\rm G}$ statistic. Our goal is to show which combinations can measure $E_{\rm G}$ to very high statistical accuracy ($< 1\%$) and provide exquisite tests of gravity at cosmological scales. For our fiducial cosmology we set the \emph{Planck} $\Lambda$CDM cosmological parameters \citep{Ade:2013zuv} and assume GR when calculating uncertainties.  

We are going to consider two modified gravity models. The first one is Chameleon gravity type models \citep{Khoury:2003aq,Brax:2004qh}, in which the scalar field which drives the cosmic acceleration couples to matter and acquires an environmentally dependent mass allowing consistence with local tests of gravity.  In these theories the $\mu$ and $\gamma$ coefficients in Equation~(\ref{eq:EGmugamma}) can be parametrised using a three-parameter set $(B_0,s,\beta_1)$ \citep{Bertschinger:2008zb}. We calculate the theoretical value of $E_{\rm G}$ for chameleon gravity using MGCAMB \citep{Hojjati:2011ix,Zhao:2008bn,Lewis:1999bs}. We are also going to consider the popular growth index parametrisation, where deviations from $\Lambda$CDM are expressed via the $\gamma_{\rm L}$ parameter and the linear growth factor can be written as $f=\Omega_m(z)^{\gamma_{\rm L}}$ \citep{Linder:2007hg}, with $\gamma_{\rm L}=6/11$ the GR value. Our chosen parameter set for the chameleon model is $(B_0,s,\beta_1)=(0.4,4,1.2)$. For the modified growth parametrisation we take $\gamma_{\rm L}=0.65$ --- note that we use $\gamma_{\rm L}$ as a trigger parameter, i.e. as a phenomenological parameter that is designed to indicate departures from GR, without the need to specify a particular theory.

As in \citet{Pullen:2015a} we quantify our results by calculating the signal-to-noise ratio (SNR) defined as
\be
{\rm SNR}^2(E_{\rm G}) = \sum_{\ell,z_i}\frac{[E^{\rm GR}_{\rm G}(z_i)]^2}{\sigma^2[E_{\rm G}(\ell,z_i)]}
\ee where $z_i$ are the foreground redshift bins. 
The $\chi^2$ value to discriminate GR and modified gravity models is written as
\be
\chi^2(E_{\rm G})=\sum_{\ell, z_i}\frac{[E^{\rm MG}_{\rm G}(\ell,z_i)-E^{\rm GR}_{\rm G}(z_i)]^2}{\sigma^2[E_{\rm G}(\ell,z_i)]},
\ee with $E^{\rm MG}_{\rm G}(\ell,z_i)$ the $E_{\rm G}$ prediction for the specific modified gravity models under consideration.

Below we present our results for the various combinations of surveys we have considered; the results are also summarised in Table~\ref{tab:results}.

\begin{table*}
\begin{center}
\caption{\label{tab:results} Forecasts of the signal-to-noise ratio (SNR) and $\chi_{\rm rms}=\sqrt{\chi^2}$ between GR and the modified gravity models under consideration for the various survey combinations we consider. For the chameleon gravity model we set $(B_0,s,\beta_1)=(0.4,4,1.2)$, while for the modified growth model we use $\gamma_{\rm L}=0.65$ (see text for further details).}  
\begin{tabular}{cccccc}
\hline
Survey &$z_c$&$z_s$&SNR&$\chi_{\rm rms}[{\rm Cham}$]&$\chi_{\rm rms}[\gamma_{\rm L}]$\\
\hline
DES $\times$ \emph{Planck} (full)&0.0--2.0&$z_{\rm cmb}$&41&4.3&1.5\\
DES $\times$ COrE-like&0.0--2.0&$z_{\rm cmb}$&85&8.9&3.0\\
LSST $\times$ \emph{Planck} (full)&0.0--2.5&$z_{\rm cmb}$&95&10.1&3.1\\
LSST $\times$ CoRE-like&0.0--2.5&$z_{\rm cmb}$&198&21.1&6.4\\
LSST $\times$ SKA\_Low-like&0.0--2.5&$z_{\rm EoR}=7$&238&25.0&8.9\\
LSST $\times$ SKA1\_Mid&0.0--2.5&$3$&47&4.8&2.1\\
LSST $\times$ SKA2\_Mid&0.0--2.5&$3$&127&12.9&5.8\\
SKA1\_Mid$^{\rm (sd)}$ $\times$  \emph{Planck} (full)&0.35--3.0&$z_{\rm cmb}$&34&3.5&1.3\\
SKA1\_Mid $\times$  \emph{Planck} (full)&0.35--3.0&$z_{\rm cmb}$&92&10.6&2.0\\
SKA1\_Mid $\times$ CoRE-like&0.35--3.0&$z_{\rm cmb}$&200&23.1&4.6\\
SKA1\_Mid $\times$ SKA\_Low-like&0.35--3.0&$z_{\rm EoR}=7$&227&25.3&6.0\\
\hline
\end{tabular}
\end{center}
\end{table*}

\subsection{Galaxy clustering $\times$ CMB lensing}

We will start by considering photometric galaxy clustering surveys (DES and LSST) combined with CMB lensing with the \emph{Planck} and CoRE-like satellites. CMB lensing is ``cleaner''  than galaxy-galaxy lensing, avoiding issues like intrinsic alignments and source redshift uncertainties. Photometric galaxy clustering surveys have been found to be able to discriminate between GR and modified gravity models more effectively than spectroscopic surveys \citep{Pullen:2015a}. That is because the RSD error, which is better measured by spectroscopic surveys, is not the dominant source of error in Equation~(\ref{eq:EGerror}). Hence, reducing the shot noise error by having higher number densities with a photometric survey is more important than precise RSD measurements. In our forecasts, we will assume a $17\%$ RSD error for DES \citep{Ross:2011zza}, {\it ie.} $\sigma(\beta)/\beta = 0.17$ in Equation~(\ref{eq:EGerror}), and $10\%$ for LSST, and consider the wavenumber range $100\leq \ell \leq 500$, since most of the signal for $E_{\rm G}$ comes from linear to quasi-linear scales for this case \citep{Pullen:2015a}. 

\begin{figure}
\includegraphics[scale=0.6]{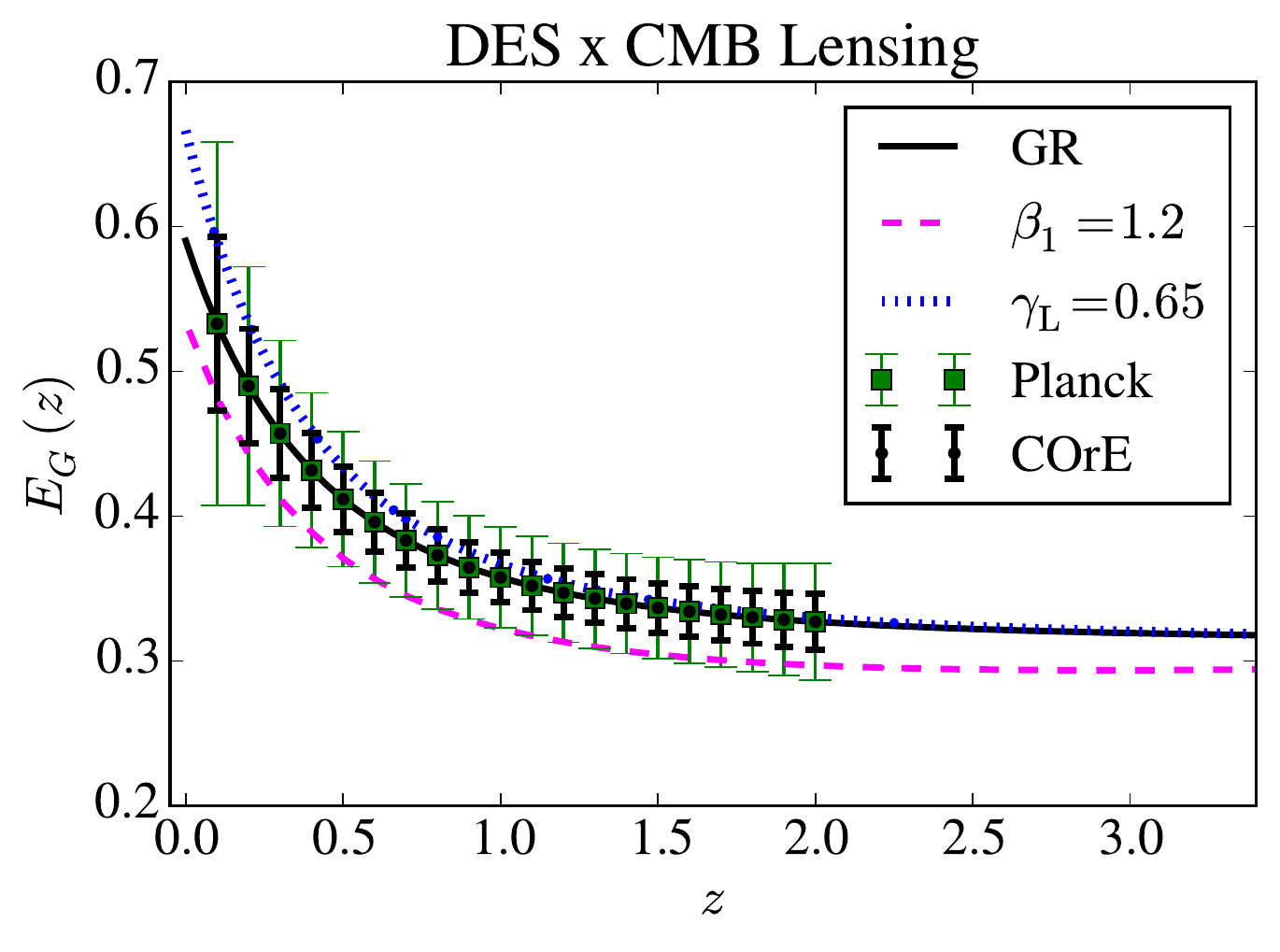}
\includegraphics[scale=0.6]{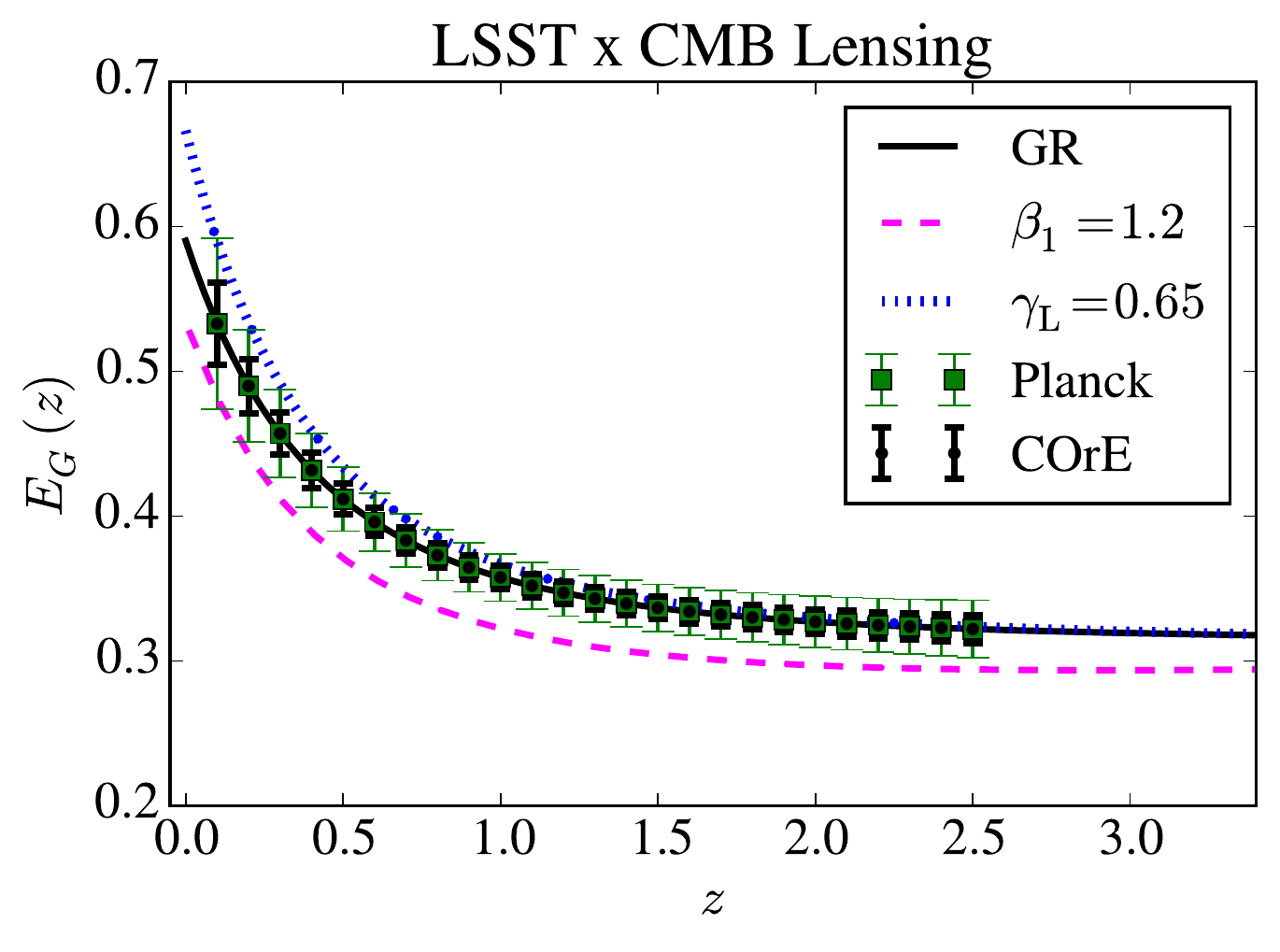}
\caption{$E_{\rm G}$ forecasts for the DES (top) and LSST (bottom) photometric optical galaxy surveys cross-correlated with the final \emph{Planck} lensing map and with the COrE-like lensing map. The Chameleon and modified growth predictions are also shown.}
\label{fig:EGforecasts1}
\end{figure}

We combine the DES and LSST galaxy clustering photometric surveys with CMB lensing measurements using the \emph{Planck} and COrE-like satellites. Our forecasts for the measurement errors are shown in Fig.~\ref{fig:EGforecasts1}.
The DES $\times$ \emph{Planck} cross-correlation gives ${\rm SNR}=41$, while DES $\times$ COrE gives ${\rm SNR}=85$. 
That is because the COrE-like satellite has a noise level more than an order of magnitude lower than \emph{Planck}. Taking LSST instead of DES we expect the results to improve significantly, as LSST covers a much bigger sky area and has increased number density of galaxies (the RSD measurement is also better). Indeed, we get ${\rm SNR}=95$ using \emph{Planck} and ${\rm SNR}=198$ using the COrE-like satellite. In the LSST $\times$ COrE case we reach fractional errors smaller than $1\%$. 
We also note that the SNR results using DES/LSST and the full \emph{Planck} lensing map provide a consistency check of our calculations as they are in agreement with the ones found in \citep{Pullen:2015a} using the same surveys. 

\subsection{Galaxy clustering $\times$ 21cm lensing}

In this Section we will first consider the combination of galaxy clustering surveys with the weak gravitational lensing of the 21cm emission from the Epoch of Reionization. This has similar advantages to using the CMB, for example precise source redshifts and the fact that we can use the full $z$-range of galaxy tracers. An additional advantage of 21cm lensing is that one is able to combine information from multiple redshift slices, which makes the lensing reconstruction noise calculated using an SKA\_Low-like instrument significantly lower than the one using the 2D CMB lensing estimator and a \emph{Planck}-like satellite \citep{Zahn:2005ap}. The possibility of measuring the lensing signal from the EoR has been studied in the past \citep{Zahn:2005ap,Metcalf:2008gq,Pourtsidou:2014pra}. Here we repeat these calculations assuming that the brightness temperature follows a Gaussian distribution, which is a reasonable approximation at the EoR, at least while the ionised regions are small. 

We consider an SKA\_Low-like instrument with collecting area $A_{\rm coll}=0.5 \, {\rm km}^2$ and maximum baseline $D_{\rm tel}=4 \, {\rm km}$ which scans half of the sky in $t_{\rm tot}=10,000 \, {\rm hrs}$. Current SKA\_Low plans include scanning a very small sky area so our chosen survey strategy is very optimistic, but we feel it is worth demonstrating the science potential of such an instrument. Another very interesting possibility if the proposed survey strategy is followed is constraining the standard cosmological parameters using 21cm lensing measurements \citep{Metcalf:2008gq}.     

In our forecasts the reionization redshift is assumed to be $z_{\rm EoR}=7$ and the observation bandwidth is $B=8 \, {\rm MHz}$.  An important issue with high redshift (low frequency) observations is the large foreground contamination. It has been shown that foreground subtraction techniques will remove the first parallel $k$ modes \citep{McQuinn:2005hk}. We model this effect by using $j_{\rm min}=4$ instead of $j_{\rm min}=1$ in Equation~(\ref{eq:Nkkell}).
Combining the above with galaxy clustering and $\beta$ measurements with LSST we find ${\rm SNR}=238$ and reach fractional errors smaller than $1\%$. That is indeed much better than \emph{Planck} and even exceeds the performance of a CoRE-like satellite. 
Considering a pessimistic case for the foreground contamination with $j_{\rm min}=10$, the lensing reconstruction noise doubles and we find ${\rm SNR}=208$, which still exceeds the performance of a CoRE-like satellite.

\begin{figure}
\includegraphics[scale=0.6]{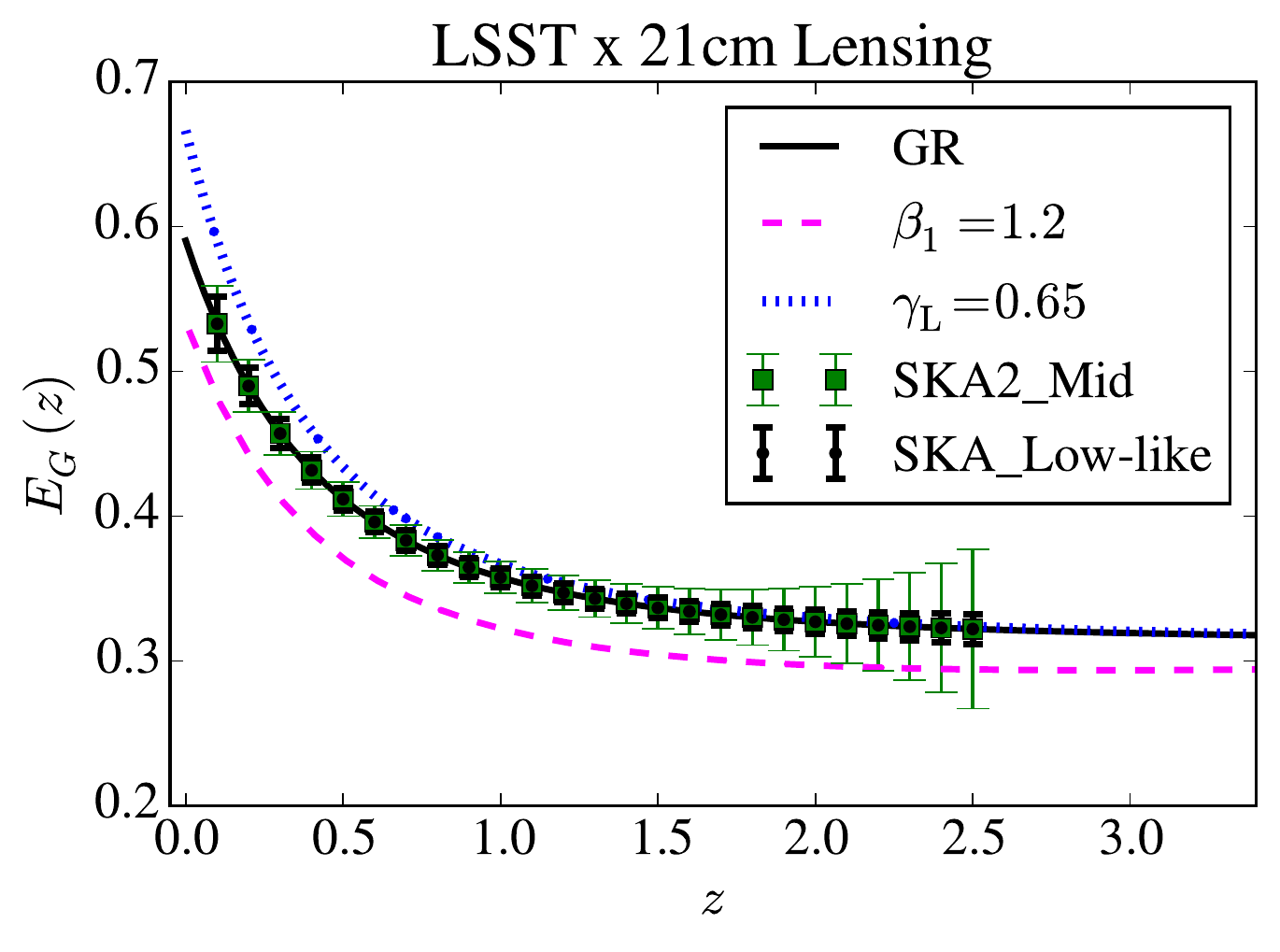}
\caption{$E_{\rm G}$ forecasts for the LSST photometric optical galaxy survey cross-correlated with 21cm lensing measurements from the EoR at redshift $z_{\rm EoR}=7$ with an SKA\_Low-like instrument, and with SKA2\_Mid 21cm lensing measurements at source redshift $z_s=3$ using the intensity mapping method. The Chameleon and modified growth predictions are also shown.}
\label{fig:EGforecasts2}
\end{figure}

We are also going to consider the lensing of 21cm emission from post-reionization redshifts (in particular $z_s=3$) probed by the SKA\_Mid array. In order to get results competitive with CMB lensing, we need to consider Phase 2 of the array --- we model the thermal noise for this case like the one of Phase 1 but one order of magnitude smaller. We also take $f_{\rm sky}=0.5$, total observation time $t_{\rm tot}=4,000 \, {\rm hrs}$ and $B=20 \, {\rm MHz}$ --- note that these numbers are realistic considering the current SKA\_Mid plans and the possibility of commissioning it to perform an intensity mapping survey. Combining with LSST we find ${\rm SNR}=127$. 
Our forecasts for the measurement errors are shown in Fig.~\ref{fig:EGforecasts2}.
Note that if we use SKA1\_Mid ({\it i.e.} Phase 1 of the array) we find ${\rm SNR}=47$.

\subsection{HI IM clustering $\times$ CMB / 21cm lensing}

In this Section we will investigate the combination of HI clustering surveys using the intensity mapping method with CMB and 21cm lensing surveys. As shown in \citet{Pourtsidou:2015mia}, there are exciting prospects for performing clustering measurements using intensity mapping surveys with the SKA and its pathfinders. Note that we will assume a $10\%$ error in the RSD parameter $\beta$ measurement, which is a very conservative estimate of the level of precision that should be achievable with an intensity mapping survey using SKA1\_Mid \citep{Raccanelli:2015hsa, Bull:2015lja}. 

We will start by considering the SKA1\_Mid  instrument in single dish (sd) mode.  An important point we need to stress is that the range of scales probed depending on the mode the instrument operates in (single dish or interferometer) is different.
Using the single dish mode we can probe very large scales, hence we are going to use $\ell_{\rm min}=10$ and $\ell_{\rm max}=2\pi D_{\rm dish}/\lambda$ for our forecasts \citep{Bull:2014rha}. For example, $\ell_{\rm max}\sim 220$ at $z\sim 1$. Using the full \emph{Planck} map we find ${\rm SNR}=34$. Since the results are not competitive with our previous forecasts, and because of 
the issue of possibly severe systematic uncertainties when probing ultra-large angular scales (like in the single-dish mode), we are going to move on to consider the interferometer mode.

\begin{figure}
\includegraphics[scale=0.6]{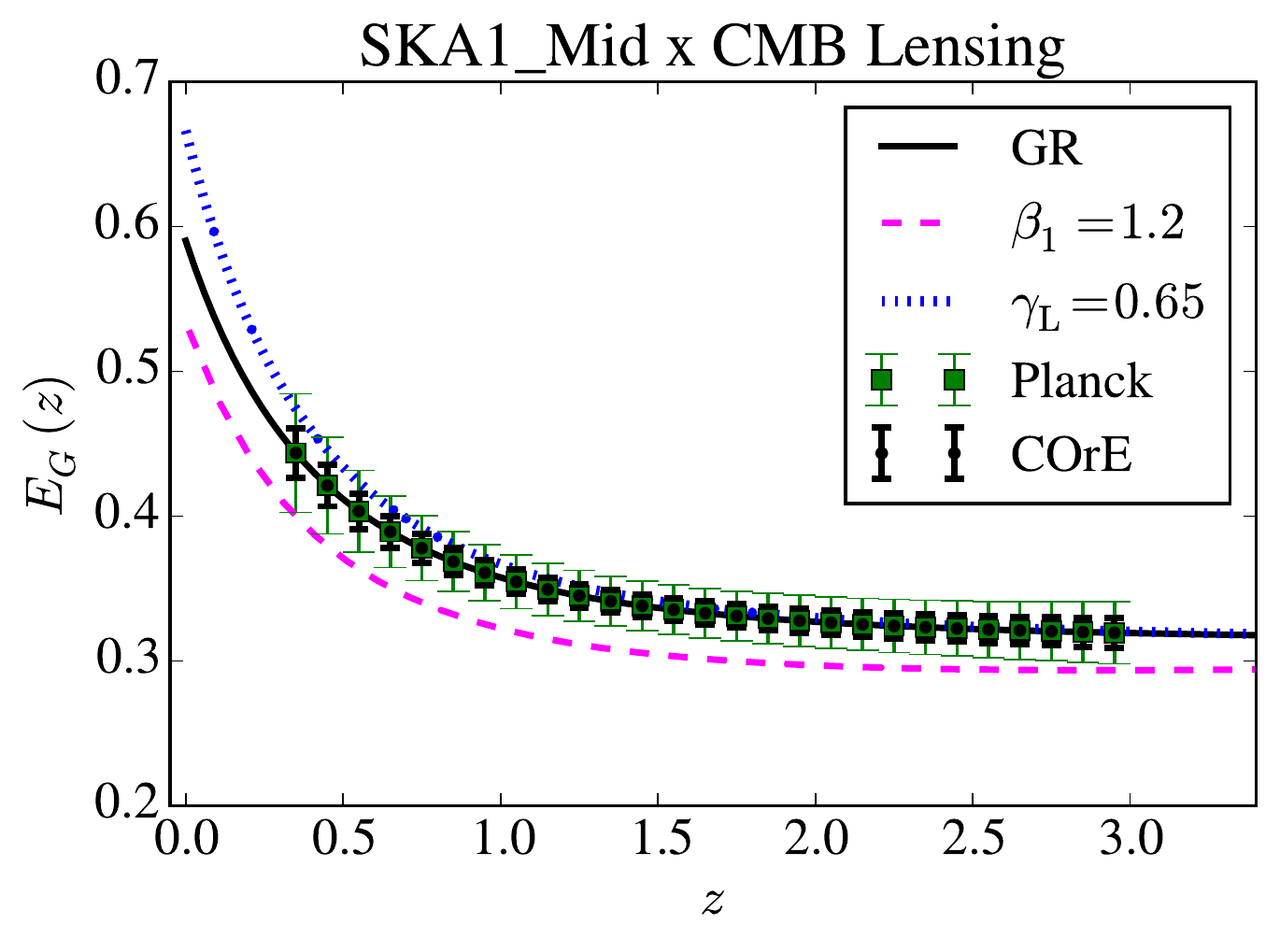} 
\includegraphics[scale=0.6]{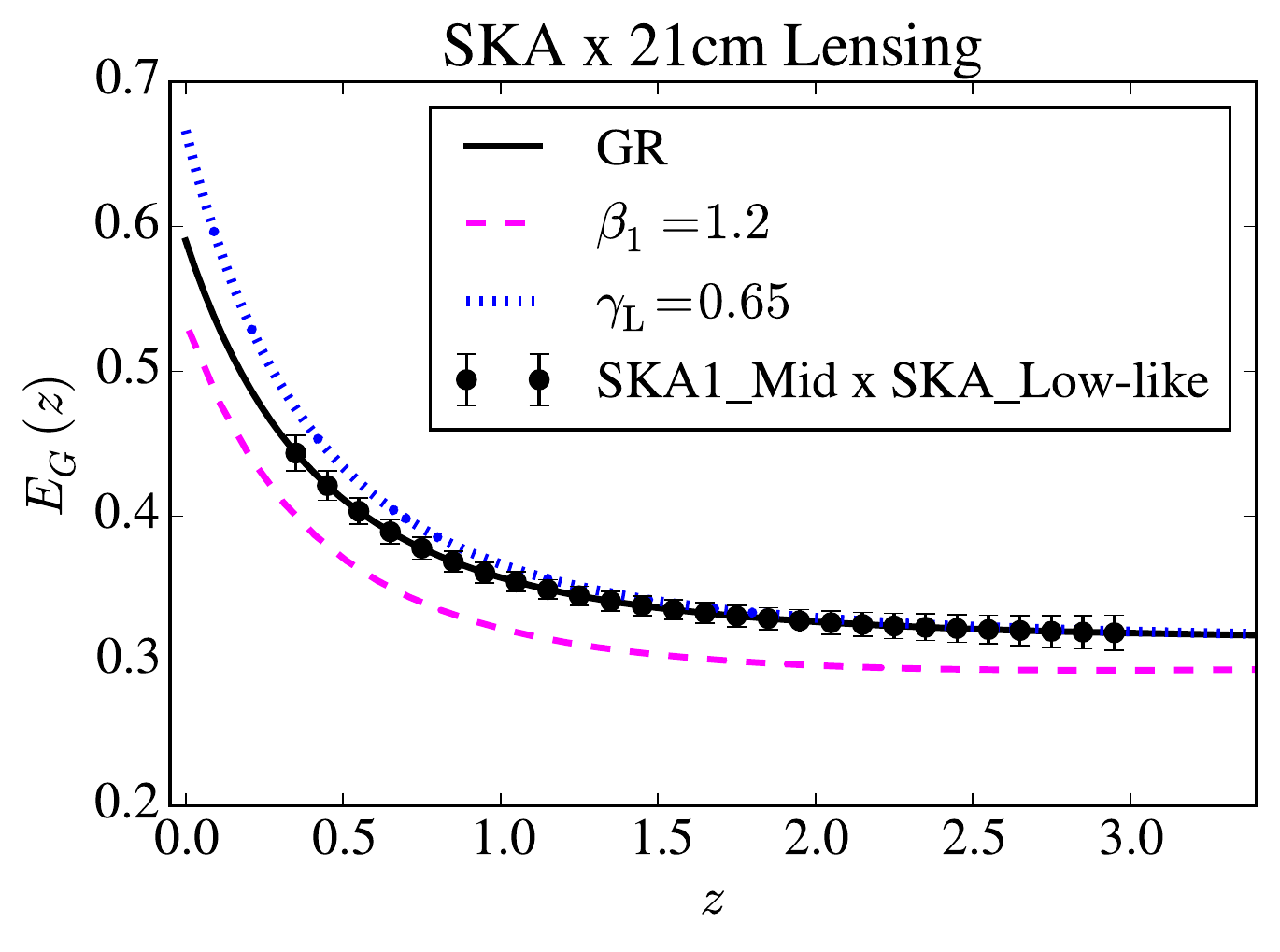}
\caption{$E_{\rm G}$ forecasts using the SKA1\_Mid instrument cross-correlated with the final \emph{Planck} lensing map and with the COrE-like lensing map (top) and the SKA\_Low-like 21cm lensing EoR map (bottom). The Chameleon and modified growth predictions are also shown.}
\label{fig:EGforecasts3}
\end{figure}

Using SKA1\_Mid in interferometer mode we get much better results. In this case we can let $\ell_{\rm max}=500$ like in the case of the optical galaxy surveys we analysed above, while the minimum multipole at each redshift is $\ell_{\rm min}=2\pi D_{\rm dish}/\lambda$. This gives $\ell_{\rm min} \sim 330$ at $z \sim 0.3$ and $\ell_{\rm min} \sim 110$ at $z \sim 3$, while $\ell_{\rm min} = 100$ for the optical galaxy surveys. Using the full \emph{Planck} map we find ${\rm SNR}=92$, while with COrE we reach ${\rm SNR}=200$, achieving fractional errors $< 1\%$ in the $E_{\rm G}$ measurements. This result implies that (assuming the problem of foreground contamination is alleviated) Phase 1 of the SKA can perform an intensity mapping survey with HI clustering measurements that are directly competitive with the galaxy clustering precision measurements by LSST. Note that considering Phase 2 of the SKA (which we model like SKA1 but with the noise level decreased by an order of magnitude) does not considerably improve the results, as the CMB lensing and $\beta$ errors dominate. Our forecasts for the measurement errors using the interferometer mode are shown in the top panel of Fig.~\ref{fig:EGforecasts3}.

Finally, we consider the case where HI clustering measurements performed with SKA1\_Mid are combined with the 21cm EoR lensing case we studied previously using an SKA\_Low-like instrument. We find ${\rm SNR}=227$, with fractional errors in the $E_{\rm G}$ measurements below $1\%$. Our forecasts for the measurement errors for this case are shown in the bottom panel of \ref{fig:EGforecasts3}.

As we have already mentioned, our signal-to-noise and $\chi_{\rm rms}$ results are summarised in Table~\ref{tab:results}. We can see that we are able to differentiate between general relativity and modified gravity 
at the level of several $\sigma$ in a number of cases. The discriminating power of the measurements we have considered is larger for the Chameleon model, as it does not converge to the GR value at high redshifts ($z>1$), while the modified growth model does. We will further comment on our results in Section~\ref{sec:conclusions}. Before we conclude, we will show the clustering and noise terms for the various surveys we have considered.

\subsection{Noise terms comparison}

To consolidate our results, we compare the noise terms used for the various survey combinations studied in this work. 

\begin{figure}
\includegraphics[scale=0.6]{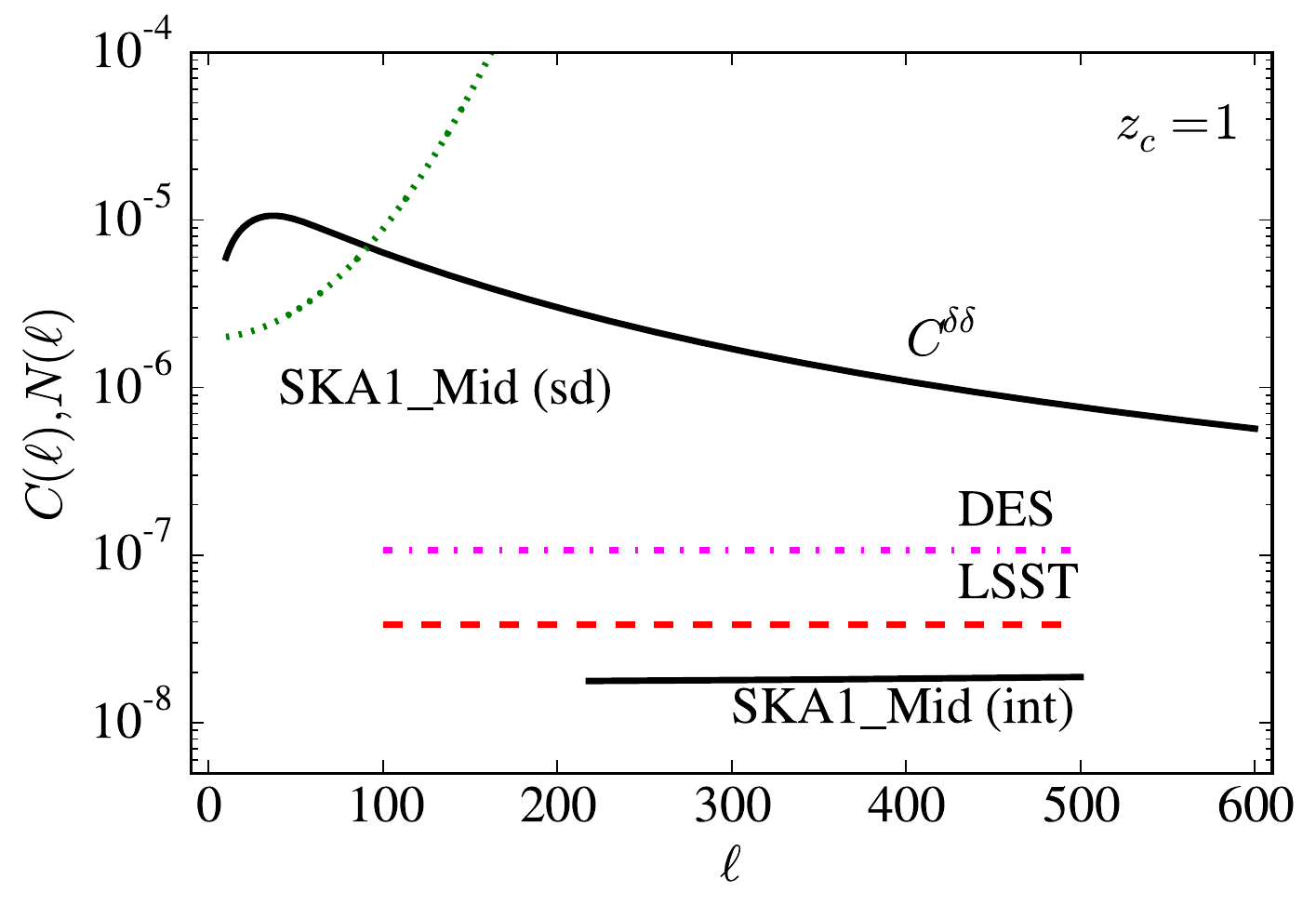}
\includegraphics[scale=0.6]{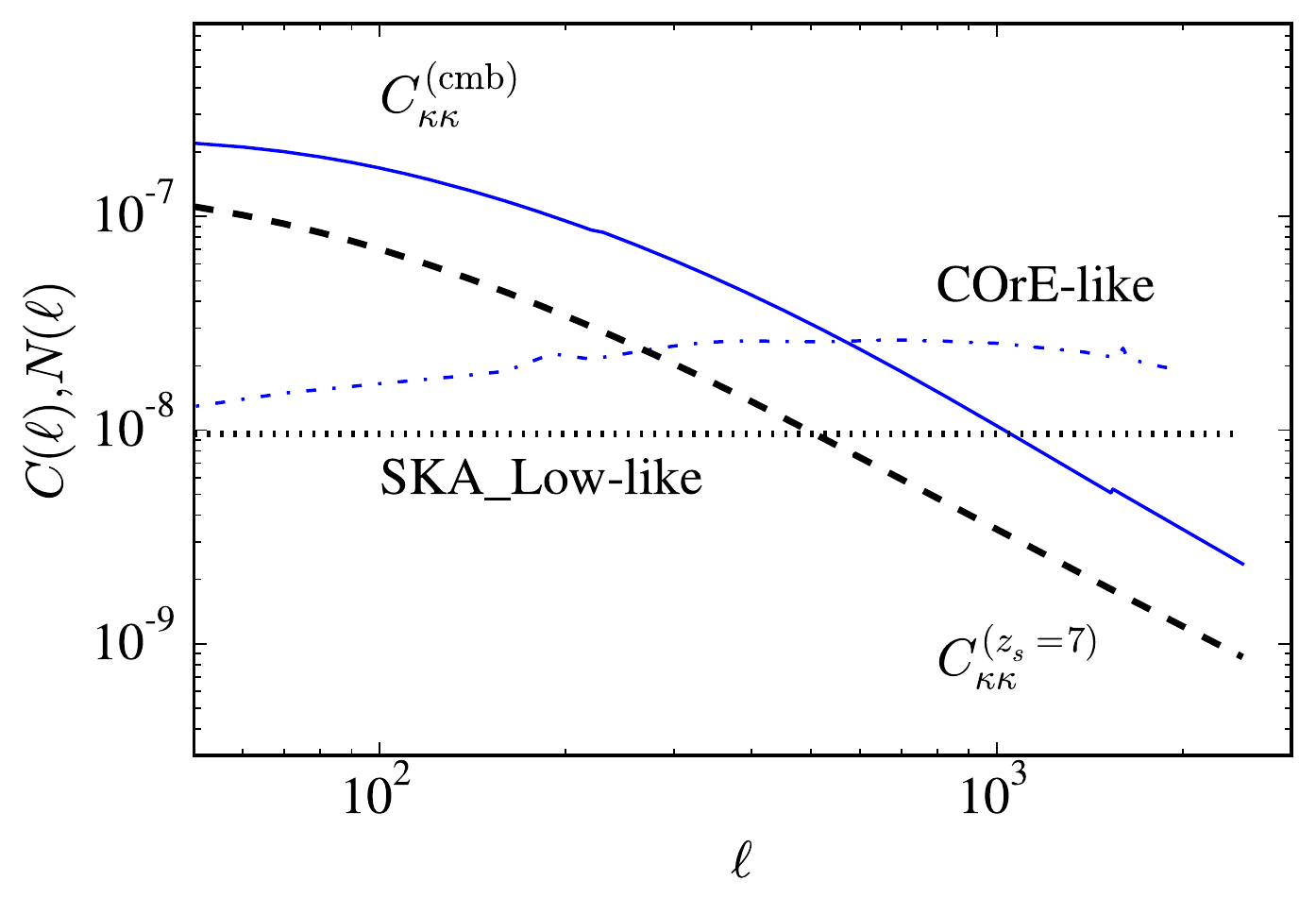}
\caption{{\it Top}: A comparison of the clustering noise terms for the various surveys considered for the bin with central redshift $z_c=1$. Note that $C^{\delta \delta} =C^{\delta_{\rm HI} \delta_{\rm HI}}/b^2_{\rm HI}=C^{\delta_g \delta_g}/b^2_g$.
{\it Bottom}: A comparison of the lensing noise terms for various surveys considered. We show the most competitive lensing measurements, {\it i.e.} using CMB lensing with a COrE-like satellite, and 21cm lensing from the EoR with an SKA\_Low-like instrument.
See text for further details.}
\label{fig:noisecomp}
\end{figure}

The top panel of Fig.~\ref{fig:noisecomp} compares the tracer density power spectra and noise terms for DES (dotted-dashed magenta line), LSST (dashed red line), SKA1\_Mid single-dish mode (dotted green line) and SKA1\_Mid interferometer mode (solid black line) for the bin with central redshift $z_c=1$. Here we note that because of the non-uniform $n(\ell)$ antennae distribution the SKA1\_Mid (int) noise curve is flat at large scales while at smaller scales (which we do not show here as they are not used) it increases as $\sim \ell^2$. Therefore, at the scales of interest for $E_{\rm G}$ the SKA1\_Mid (int) instrument has its minimum thermal noise value; however, the minimum $\ell$-scale it can probe is larger than the one of the optical galaxy surveys considered.  
The solid black curve is the angular power spectrum $C^{\delta \delta} =C^{\delta_{\rm HI} \delta_{\rm HI}}/b^2_{\rm HI}=C^{\delta_g \delta_g}/b^2_g$ at $z_c=1$. 

The bottom panel of Fig.~\ref{fig:noisecomp} compares the lensing convergence power spectra and noise terms for the COrE-like satellite (blue dotted-dashed line) and the SKA\_Low-like interferometer (dashed black line). As we have already mentioned, the {\emph Planck} noise is approximately one order of magnitude higher than COrE, while the SKA2\_Mid noise level for HI sources at $z_s=3$ is about three times higher than COrE (and, of course, the lensing convergence power spectrum is also lower at lower redshifts).

\section{Discussion and conclusions}
\label{sec:conclusions}

In this work we considered HI intensity mapping clustering and lensing as probes of the clustering bias-free $E_{\rm G}$ statistic which can be used to test general relativity on cosmological scales. We forecasted the ability of various survey combinations -including intensity mapping, (photometric) optical galaxy and CMB lensing surveys- to test GR and constrain modified gravity theories, in particular Chameleon gravity and modified growth index parametrisation models. 

Our results show that fractional errors $< 1\%$ in the $E_{\rm G}$ measurement can be achieved in a number of cases, namely SKA1\_Mid (LSST) HI intensity mapping (galaxy) clustering combined with COrE CMB lensing or, alternatively, with a SKA\_Low-like EoR 21cm lensing survey. Looking at the corresponding $\chi_{\rm rms}$ values in Table~\ref{tab:results}, which are well above unity, we see that these surveys can  provide strong constraints on the modified gravity models parameters. The modified growth index model is more difficult to constrain as its expectation $E_{\rm G}$ value becomes indistinguishable from GR at $z>1$ --- that is why the $\chi_{\rm mrs}(\gamma_{\rm L})$ values are larger when using LSST (instead of SKA1\_Mid), as LSST covers the $z<0.35$ range where the differences between the growth parametrisation model and GR are more pronounced (the optical galaxy surveys also cover a bigger $\ell$-range than SKA1\_Mid in interferometer mode). It would therefore be very beneficial if the SKA1\_Mid clustering and RSD measurements were extended to include the low redshift regime, $0<z<0.35$ (Band 2).

The $E_{\rm G}$ statistic is clustering bias free, but its statistical error is affected by the bias and it is useful to comment on how sensitive it is to changes in the analytical formulae we have used. For example, considering a factor of $2$ smaller bias (i.e. $\sqrt{1+z}/2$ for the galaxy case and the analogous expression for the HI case), we still get less than a percent fractional errors in our most competitive cases and the total signal-to-noise ratio is reduced by about $3\%$ (on the other hand, a larger bias increases the SNR very slightly). That is because a smaller (larger) bias increases (decreases) the contribution of the shot / thermal noise terms. 

In terms of the required precision in the RSD measurements ($\sim 10\%$), the intensity mapping method appears to be quite advantageous. We already mentioned that in order to produce competitive $E_{\rm G}$ measurements using optical galaxy surveys one needs to use photometric instead of spectroscopic surveys, but measuring RSD this way is difficult \citep{Pullen:2015a}. On the other hand, using intensity mapping and the SKA excellent redshift information is automatically provided and one can get results competitive with the ones by upcoming spectroscopic surveys like Euclid \citep{Raccanelli:2015hsa, Bull:2015lja}. Furthermore, a very recent paper on measuring $E_{\rm G}$ using number counts \citep{Dizgah:2016bgm}  
showed that the usually neglected lensing contribution to galaxy number counts (which affects the $gg$ and $g\kappa$ spectra) is important -especially at high redshifts- and renders $E_{\rm G}$ bias dependent. Intensity mapping does not suffer from this problem, because there is no magnification term at linear order as surface brightness is conserved. 

Another feature of modified gravity (and exotic dark energy) theories can be scale dependence. For example, $E_{\rm G}$ is found to be strongly scale dependent in the case of $f(R)$ gravity theories \citep{Pullen:2015a}, so one could also use $E_{\rm G}(k)$ measurements to constrain scale dependent gravity. In this work we have mild scale dependence only in the case of Chameleon gravity, so we just averaged over the wavenumber range at each redshift bin for our predictions. However, this is a potentially very interesting subject and we plan to investigate it in future work.       

An important point we need to stress is the need to control systematics. Future measurements will reach an unprecedented level of statistical precision ($< 1\%$ ) and if systematic effects are not correctly identified and removed the total error will be much larger. Details about systematics when combining galaxy surveys with CMB lensing can be found in \citet{Pullen:2015b}. 
In HI intensity mapping clustering and lensing surveys the biggest problem is the presence of galactic and extragalactic foregrounds.
These can be orders of magnitude brighter than the HI signal but they have a smooth, power-law frequency dependence, in contrast to the fluctuating signal, so they can be removed \citep{Morales:aa, LiuTegmark, Alonso15}. In order to identify systematics and test the various foreground removal techniques it is essential to perform auto- and cross-correlation clustering and lensing studies using intensity mapping and optical galaxy surveys. For this purpose we can exploit SKA pathfinders like MeerKAT \citep{Pourtsidou:2015mia}. These studies will also give us precise measurements of the mean HI brightness temperature $\bar{T}(z)$, which is assumed to be known in our forecasts. 

To conclude, the intensity mapping technique, although still in its infancy, is in principle ideal for testing general relativity and the standard cosmological model on large scales. Intensity mapping surveys performed with the Square Kilometre Array have the advantage of excellent redshift resolution and they can map a large fraction of the sky across a wide range of scales and redshift, achieving very high signal-to-noise measurements. At large scales in particular, an intensity mapping survey with SKA1 can be directly competitive with state-of-the art photometric optical galaxy surveys like LSST. Combining this with CMB lensing using COrE-like satellites or 21cm lensing from the EoR with an SKA\_Low-like array we can perform exquisite tests of gravity and, consequently, help unravel the secrets of the dark sector of the Universe. 

 \section*{Acknowledgments}
This work was supported by STFC grant ST/H002774/1. I would like to thank David Bacon, Davide Bianchi, Robert Crittenden, Dida Markovi\v{c} and Matthew Withers for useful discussions.


\bibliographystyle{mn2e}
\bibliography{references_EG}

\end{document}